# Title: Fractal Universality in Near-Threshold Magnetic Lanthanide Dimers

**Authors:**

Constantinos Makrides[1,2], Ming Li[1], Eite Tiesinga[2], Svetlana Kotochigova[1*]

**Affliliations:**

[1]Department of Physics, Temple University, Philadelphia, PA 19122-6082, USA

[2]Joint Quantum Institute and Joint Center for Quantum Information and Computer Science, National Institute of Standards and Technology and University of Maryland, Gaithersburg MD 20899, USA

[*]Correspondence to: skotoch@temple.edu

**Abstract:** Ergodic quantum systems are often quite alike, whereas nonergodic, fractal systems are unique and display characteristic properties. We explore one of these fractal systems, weakly bound dysprosium lanthanide molecules, in an external magnetic field. As recently shown, colliding ultracold magnetic dysprosium atoms display a soft chaotic behavior with a small degree of disorder. We broaden this classification by investigating the generalized inverse participation ratio and fractal dimensions for large sets of molecular wave functions. Our exact close-coupling simulations reveal a dynamic phase transition from partially localized states to totally delocalized states and universality in its distribution by increasing the magnetic field strength to only a hundred Gauss (or 10 mT). Finally, we prove the existence of nonergodic delocalized phase in the system and explain the violation of ergodicity by strong coupling between near-threshold molecular states and the nearby continuum.

**Introduction** Ultracold atomic physics is now poised to enter a new regime, where far-more complex atomic species can be cooled and studied. Magnetic lanthanide atoms with their large magnetic moment and large orbital angular momentum are extreme examples of such species. In fact, ultracold gases of magnetic lanthanides provide the opportunity to examine strongly correlated matter, creating a platform to explore long-range dipolar magnetic interactions and exotic many-body phases such as quantum ferrofluids, quantum liquid crystals, and supersolids. In particular, experimental advances in trapping and cooling Dy, Ho, Er, and Tm atoms (*1-8*) are paving the way towards these goals.

Most lanthanides have a partially-filled submerged 4f-electron shell, which lies beneath a filled $6s^2$ shell. The 4f-electrons are added with parallel spins leading to both a large magnetic moment and a large total and orbital angular momentum. These "unquenched" angular momenta generate a rich atomic and molecular structure as well as collective phenomena of these systems (*8,9*)

In previous studies (*8-11*) we explored the scattering and interactions between neutral magnetic lanthanide atoms in the ultracold regime with temperatures well below 1 mK. Over the last few years we developed a framework for understanding the complex anisotropic interactions between these dipolar atoms and their chaotic behavior. In particular, our theoretical model in combination with advanced numerical treatments bridges the conceptual gap between "simple" alkali-metal and alkaline-earth atoms and complex lanthanides. It allows us to explain the origin of the dense spectra and chaotic character found in the statistics of the observed Er and Dy collisional resonances as due to the anisotropy of both the short- and long-range interactions between the atoms.

In this paper, we apply modern theoretical approaches from quantum theories of chaos to characterize the observed chaotic distribution of Fano-Feshbach resonances in weak magnetic fields. We turn to an analysis of near-threshold bound states and, in particular, the eigenstate wavefunctions of our molecular Hamiltonian matrices, which include both localized and continuum basis states or configurations. By introducing generalized moments of wave function mixing coefficients we are able to identify extended eigenfunctions as well as a number of localized eigenstates in our system. Localized states are characterized by a wave function with a small percentage of mixing. Our extended eigenfunctions exhibit strong, disordered mixing that is partially chaotic and approaches a unique type of *universal* behavior. The degree of mixing is controlled by the external magnetic field $B$. We compute the *fractal* statistics of the extended states and attempt to answer the question, whether our quantum system has ergodic properties where all accessible configurations are equally available?

Our model goes beyond the canonical theory of spectral statistics. It focusses on a discrete set of bound rovibrational levels coupled to scattering continua, where states exist for any energy. The level statistics is only chaotic in the energy region of the bound states. In practice, we treat continua with a discretization procedure by introducing a large spatial/radial cutoff. As the result, we include a large number of continuum states in order to converge the eigenenergies and eigenvectors of the selected bound states. This approximation works well for many scattering problems.

Other types of ultracold atomic and molecular physics experiments are also starting to influence the developments in the quantum chaos and disorder theories. This is mainly due to the fact that such experiments have an unprecedented control and tunability of system parameters. Disorder potentials, for example, can be used to study Anderson localization (*12*). For ultracold atoms and molecules these potentials can be created with speckle patterns, obtained by passing laser light through glass plates that introduce random phase profiles (*13,14*) or by preparing randomly located dipoles in optical lattices, which are periodic potentials for atoms and molecules (*15,16*). In the former case the strength of the disorder potential is determined by the laser intensity. In addition, atom-atom interactions can be set at a known value with collisional Fano-Feshbach resonances (*17*), which is then seen to modify the localization (*18*). For randomly distributed dipoles the eigenenergy spectra are controlled by the lattice filling fraction.

Disordered quantum systems exhibit a rich diversity of statistical properties and types of extended and localized eigenstates. Within an early statistical classification scheme they were often characterized by two distinct regimes where the nearest-neighbor spacing distribution (NNSD) of eigenvalues follows either an integrable Poissonian or chaotic Wigner-Dyson statistics (*19*). At the same time, there exist many dynamical systems that are partially chaotic (*20,21*). In this intermediate regime, the best fit to NNSD is obtained using Brody distribution (22). Recent studies of Anderson localization on hierarchical lattices such as random regular graphs (RRG) point to the existence of a phase transition that separates extended-ergodic (EE) states from non-ergodic, multifractal extended (NEE) states (*16,23-25*). The correspondence between classical and quantum chaos in ultracold dipolar collisions is examined in Ref. (*26*).

Here, we broaden the class of partially chaotic systems accessible for theoretical and experimental study. We present the search for fractal dimensions in disordered scattering and near-threshold bound states of ultracold highly-anisotropic bosonic and magnetic $^{164}$Dy lanthanide atoms in an external magnetic field. Their high-density Fano-Feshbach resonance spectrum was recently seen in a number of experiments with dysprosium (*8,27,28*) and is without precedent in other ultracold quantum gases. For comparison, the resonance density in alkali-metal atom collisions is two orders of magnitude lower.

We concentrate on the hundreds of loosely-bound dysprosium molecular states, with binding energies less than 0.5 GHz when expressed in units of the Planck constant $h$, as we increase the strength of an external magnetic field $B$ from zero to 250 G, where 1 G equals 0.1 mT. Unique is the notion of locality and the role of scattering thresholds. In the Anderson Hamiltonian with random onsite energies this is defined through the location of the particle or spin. Eigenstates are superpositions of being at different locations. In our system, the zero-$B$-field molecular eigenstates are the natural equivalent (local) states. These $B = 0$ states have energies either below or above the thresholds. Finite $B$-field eigenstates are then superpositions of zero field eigenstates. This is schematically illustrated in Fig 1. The $B = 0$ Hamiltonian matrix (Fig. 1a) is diagonal with values that are randomly distributed. For non-zero magnetic field off-diagonal matrix elements appear due to the Zeeman interaction and are shown as yellow boxes in Fig. 1b. Our Hamiltonian matrix has a few hundred bands.

By focusing on the mixing coefficients of molecular wave functions we find a transition between two non-ergodic extended (NEE) regimes as we increase the applied magnetic field. We also isolate localized states within our chaotic level structure. The transition is further elucidated by computing the spectrum of multifractality using concepts of Ref. (*23*) and correlated with statistical descriptions of the nearest-neighbor energy spacing distributions of the near-threshold quantum states. Our previous analyses (*8,9*) have already showed that the spacing distribution of these resonances closely, but not completely, follows a Wigner-Dyson distribution for the larger magnetic field strengths.

**Results**: The Hamiltonian of our diatomic $^{164}$Dy quantum system in a homogeneous magnetic field with strength $B$ was discussed in Ref. (*8*) and only a brief account of the features relevant in this discussion are presented here. The atoms have angular momentum $\vec{j}_i$ with $i = a$ or $b$, $j_i = 8$, and projection $m_i$ along the magnetic field. For $B = 0$ the total molecular angular momentum $\vec{J} = \vec{\ell} + \vec{j}_a + \vec{j}_b$ and parity, i.e. even versus odd relative orbital angular momentum $\vec{\ell}$ or partial wave, are conserved. For even parity the Hamiltonian describes rovibrational motion in 81 distinct *gerade* anisotropic potentials that depend on the inter-atomic separation and atomic spins. A schematic of the molecular potentials, scattering thresholds, and a rough estimate of the level density in the relevant energy region is shown in Fig. 2. A sensible choice of basis functions is the weakly-bound and above-threshold continuum $B = 0$ eigenstates $|vJM\rangle$ with eigenvalue $E_v^{JM}$. For non-zero field and fixed $M$, unit-normalized eigenfunctions are then

$$|\Psi; M\rangle = \sum_{J \geq |M|} \sum_v c_{vJM} |vJM\rangle \quad (1)$$

with expansion coefficients $c_{vJM}$. The coefficients are determined by diagonalizing the Hamiltonian with diagonal matrix elements $E_v^{JM}$ coupled (and shifted) by the matrix elements of the Zeeman Hamiltonian. Crucially, the coupling strengths are proportional to B and only non-zero when $|\Delta J| = 0,1$. Hence, the Hamiltonian leads to a banded matrix that is schematically depicted in Fig 1. Figure 3 shows the exact eigenenergies of our Hamiltonian for four typical B-field strengths. The numerical approach and convergence properties of the Hamiltonian are described in Materials and Methods.

We analyze the fractal properties of eigenfunctions of near-threshold bound states of $^{164}$Dy$_2$ that can lead to the magnetic Fano-Feshbach resonances, when their binding energy goes to zero for increasing $B$. Our analysis follows the spectral statistics and the spectrum of fractal dimensions (SFD) as used by Ref. (*23*). We define the generalized moment or *partition function* as well as a *grand potential* for each eigenfunction $|\Psi; M\rangle$ as

$$I(q) = \sum_{i=1}^{N}|c_i|^{2q} \text{ and } \tau(q) = -\log_N I(q), \quad (2)$$

respectively, where $q \in \mathbb{R}$ is a generalization of the inverse of a temperature and, for simplicity, we suppressed the eigenstate labels $\Psi$ and $M$ and replaced the indices $vJM$ by the single index $i$. Here, $\log_N$ is the base-$N$ logarithm. Unit normalization of the eigenstates ensures that $I(q = 1) = 1$ and $\tau(q = 1) = 0$. The function $I(q)$ can also be interpreted as a generalization of the inverse participation ratio (IPR) to which it coincides for $q = 2$.

The functions $I(q)$ and $\tau(q)$ classify eigenstates of the disordered Hamiltonian. For example, a localized eigenstate with a single dominant $c_i$ has $I(q) \approx 1$ and $\tau(q) \approx 0$ independent of $q$ and $N$. On the other hand, near equal population with $|c_i|^2 \approx 1/N$ for all basis states $i$ leads to $I(q) \approx N^{1-q}$ and $\tau(q) \approx q - 1$. Hence, the *grand potential* $\tau(q)$ does not depend on the number of basis functions. In fact, the set of eigenstates $|\Psi; M\rangle$ *within* a small eigenenergy interval that still contains a fairly large number of eigenstates is *ergodic* when the "local", i.e. fixed $i$, average $\langle |c_i|^{2q} \rangle$ over such a subset of eigenstates approaches $\langle I(q) \rangle / N$ when $N \to \infty$. The system is then said to uniformly explore all basis functions with relatively weak fluctuations. On the other hand, a departure from these relations and, in particular, $\tau(q) \neq q - 1$ is a signature of *nonergodic* eigenstates.

Figures 4 (A to D) show the *grand potential* $\tau(q)$ for Dy$_2$ eigenstates $|\Psi; M\rangle$ with eigenenergy in $[\mathcal{E}, 0]$ at four finite $B$-field values and $\mathcal{E}/h = -0.5$ GHz. All curves are monotonic and concave, meet at $q = 0$ and 1, and for $q > 1$ approach a linear functional form with a slope that is less than one. The curves can be compared to $\tau(q) = 0$ for a localized state (or any of our eigenstates at $B = 0$) and $\tau(q) = q - 1$ for ergodic GOE states. Here, GOE is an abbreviation for the Gaussian orthogonal ensemble of N-dimensional Hamiltonians, whose matrix elements are randomly chosen from a Gaussian distribution. Averages over eigenstates within a small energy window for the whole ensemble of these GOE Hamiltonians satisfy $\tau(q) \to q - 1$ for $N \to \infty$. In the context of metal-to-insulator transitions (*30*), $\langle \tau(q) \rangle$ has been used to distinguish insulator phases with $\langle \tau(q) \rangle = 0$ from metallic states with $\langle \tau(q) \rangle \neq 0$. Figure 4 (A to D) also show that $\tau(q)$ not only depends on near-threshold eigenstates, but also magnetic field. At $B = 10$ G with already significant Zeeman couplings between zero-field basis states there remains a relatively large number of localized states with slope close to zero for $q > 1$. For stronger $B$ fields mixing increases and more and more eigenstates have a finite slope. At $B = 250$ G only one state remains localized, while $\tau(q)$ for all other states seem to collapse onto a single universal curve with a critical slope that is less than one for $q > 1$. We identify this as the unique universal characteristic of disorder in our system.

We further quantify these observations by studying the fractal dimension $D_q = \langle\langle \tau(q) - \tau(1) \rangle\rangle/(q-1)$. Here $\langle\langle \cdots \rangle\rangle$ indicates a double average over eigenfunctions with energy in $[\mathcal{E}, 0]$ and over realization of the molecular Hamiltonian. In particular, we changed the strength of the strongest anisotropic potential contribution, as identified in Ref. (8) in the zero-field Hamiltonian over ±5% from its physical strength assuming a uniform probability distribution. This range of strengths is sufficiently large that the $B = 0$ bound state spectrum changes significantly but at the same time is small enough to not change its mean spacing and random distribution. (A ±5% variation also reflects the uncertainty in this strength.) As a consequence, the accuracy of our statistical analysis improves without changing our conclusions.

Figure 4E shows the computed $D_q$ as a function of $B$ for $q = 1$ and 2. For $q \to 1$, $D_q$ corresponds to the derivative of averaged $\tau(q)$ with respect to $q$ and, in fact, is an effective entropy

$$S = \lim_{q \to 1} D_q = \langle\langle -\sum_{i=1}^{N} |c_i|^2 \log_N(|c_i|^2) \rangle\rangle \quad (3)$$

of the probabilities $|c_i|^2$. By construction the entropy $0 \leq S \leq 1$ and only equals one when $|c_i|^2 = 1/N$ for all $i$. By construction, the entropy is $0 \leq S \leq 1$ and $S$ only equals one when $|c_i|^2 = 1/N$ for all $i$. The choice of $D_{q=2}$ is a compromise. It gives a reasonable estimate of the large $q$ trends while at the same time it describes the dimensionality near the IPR point.

Two observations can be made about Fig. 4e. Firstly, we note that the fractal dimension $D_q$ depend on q. This is characteristic of a multifractal system. Secondly, at fixed $q$ the entropy S and $D_2$ linearly increase with magnetic field starting from zero, i.e. the system becomes chaotic, and then saturates at a finite value near 0.5 much smaller than one. The transition between the two behaviors occurs between 50 G and 100 G. Figure 4e also shows the square root of the variance of the fractal dimension as a function of $B$. A smaller variance indicates increased universality with a larger fraction of the eigenfunctions with a similar $\tau(q)$. It is smallest for $B > 200$ G.

A measure of the multifractality of $D_q$ or equivalently the $q$-dependence of $\langle\langle\tau(q)\rangle\rangle$ is given by the spectrum of fractal dimension (SFD) or the Legendre transform of $\langle\langle\tau(q)\rangle\rangle$ (*30*). It is defined as $f(\alpha) = min_{q \in \mathbb{R}}[q\alpha - \langle\langle\tau(q)\rangle\rangle]$ and implies $d\langle\langle\tau(q)\rangle\rangle/dq = \alpha$. Figure 5a shows $f(\alpha)$ as functions of $\alpha$ for six representative $B$ values ranging from 1 G to 250 G. Each curve is defined between $(\alpha_-, \alpha_+)$ with $0 < \alpha_- < \alpha_+$. They are positive and resemble half of an ellipse with a maximum value $f_{max} \approx 1$ at $\alpha = \alpha_{max}$, the slope of $\tau(q)$ at $q = 0$. A small value of $f(\alpha)$ indicates that corresponding tangent lines of $\tau(q)$ extrapolate to the origin. Finally, $\alpha_-$ is the slope of $\tau(q)$ when $q \to +\infty$.

We characterize the behavior of $f(\alpha)$ as *multifractal* with a trend towards the *ergodic* limit for increasing magnetic field strength, where the ergodic limit corresponds to a delta function at $\alpha = 1$. For increasing $B$ the SFD sharpens up with a maximum that shifts to smaller $\alpha$ and approaches $\alpha_{max} = 2.3$. The approach is shown in Fig. 5b. The SFD reaches a universal function for $B > 50$ G leading to the conclusion that weak *multifractality* persists over the whole range of magnetic fields accessible in our numerical analyses. In contrast to results of Ref. (*23*) for a Bethe lattice, we do not obtain a triangular shaped SFD even for small magnetic field strengths. The shape of our curves is always flat topped typical for weak *multifractality*. The lack of *ergodicity* in the collisional dynamics of Dy atoms will have significant implications for the observed spectra and invites further investigation.

The onset to universality has its origin in our Hamiltonian matrix structure. It contains the $B=0$ G eigenenergies on the diagonal with mixing and shifts induced by the Zeeman Hamiltonian with each of its matrix elements proportional to $B$. Naively, we can then expect that the expansion coefficients of eigenstates $|\Psi, M\rangle$ become independent of $B$ when the energy scale of the Zeeman interaction is larger than that of the $B=0$ G eigenenergies. In practice, however, the energy scales for the two contributions are hard to estimate, because of the presence of the (discretized) continua near our threshold bound states with energy between $[\mathcal{E}, 0]$. The coupling strength between bound and spatially-extended continuum states has a different character than those between bound $B=0$ G basis states. In fact, we have only been able to empirically determine this $B$ value (that is, between 50 G and 100 G) from the spectrum of the fractal dimensions.

We finish by connecting the statistical properties of eigenstate wavefunctions with those of the NNSD of their eigenenergies. This spectral statistics of quantum systems is often characterized by either a Poisson distribution $P_P(\Delta) = e^{-\Delta}$ or a Wigner-Dyson distribution $P_{WD}(\Delta) = (\pi/2)\Delta e^{-\pi \Delta^2/4}$ (30), where $\Delta = s/\langle s \rangle$ is the dimensionless scaled spacing between levels. The Brody distribution is used to describe quantum systems that experience a transition from Poisson to Wigner-Dyson statistics (21). Here, we construct the NNSD as a histogram from eigenenergy spacings within the energy interval $[\mathcal{E}, 0]$ and over several realizations of the molecular Hamiltonian, as described in the previous subsection. We then determine the Brody parameter $\eta$ that provides the best fit of the Brody distribution $P(\Delta; \eta) = b(1+\eta)\Delta^{\eta} e^{-b\Delta^{\eta+1}}$ (22) to the histograms, where $b$ is a known constant such that $\int_0^\infty \Delta^k P(\Delta; \eta) d\Delta = 1$ for $k = 0$ and 1.

Figure 6 shows our Brody parameter $\eta$ as a function of B between 0 G and 250 G. Examples of our fitting of the Brody parameter to our histograms for two $B$ fields, 25 G and 160 G, are presented as insets. We observe that the Brody parameter changes monotonically from $\eta \approx 0.1$ for $B \geq 1$ G to a $\eta$ just below 1 for $B > 100$ G. Hence, our system undergoes a transition from an integrable system with Poissonian distributed levels to a "hard-chaotic" system that satisfies the Wigner-Dyson distribution. Crucially, we observe two physical regimes: For $B < 50$ G the Brody parameter increases linearly, whereas for larger field strengths, it saturates to an asymptotic value and becomes universal. The critical or transition $B$ value is ~60 G, as defined in the figure, and closely follows the universality as observed in the fractal dimension and multifractality of Figs. 4 and 5.

**Discussion:** Here, we obtained a detailed understanding of the chaotic quantum dynamics of two weakly-bound bosonic lanthanide $^{164}$Dy atoms in a magnetic field. The near-threshold chaotic molecular structure is a consequence of the magnetic Zeeman-interaction-induced mixing of 81 *gerade* electronic potentials of $Dy_2$ that dissociate to two ground state Dy atoms (*31*). The remarkably complex nature of the electronic structure is due to the large total and orbital angular momentum of dysprosium atom.

Our work was motivated by successful experiments in producing quantum degenerate gases of dysprosium atoms (*3,8*) as well as the measurement of its dense magnetic Fano-Feshbach spectrum. The latter is directly related to the collisional properties of atoms in an optical trap. In a previous study of chaos in this system (8) we showed that the nearest-neighbor spacing distribution of Fano-Feshbach resonances was closer to the Wigner-Dyson than the Poisson distribution.

Here, we identified the key parameter that governs chaos and randomness of the vibrational diatomic wavefunctions and how it changes with increasing magnetic field to a system with a higher level of chaoticity. We analyzed the fractal properties of the Dy$_2$ eigenfunctions in the near-threshold region, which allowed us to characterize the degree of localization and ergodicity for each eigenstate. We were surprised to find that some states in the chaotic system are localized whereas others are non-localized, making our system multifractal. Moreover, by increasing the magnetic field strength we are able to change our system from partially chaotic to completely chaotic with a remarkable degree of universality. At the same time, we realized that the degree of ergodicity, characterized by fractal dimensions $D_q$, has a limiting value of ≈0.5 for large $B$, where complete ergodicity corresponds to $D_q = 1$.

**Materials and Methods:** Our Hamiltonian contains bound states as well as a continuum spectrum separated by a threshold at energy $E = 0$ corresponding to two ground-state Dy atoms at rest and in their energetically-lowest magnetic Zeeman level. Following the study of Ref. (8), it suffices to study even parity states with $M = -16$, limit the sum over $J$ to $\leq J_{max}$ and use $B = 0$ G eigenstates with $E_v^{JM} \in [E_{min}, E_{max}]$, where $E_{min} < 0 < E_{max}$. The continuum spectrum is "discretized" by limiting the extend of the vibrational motion to a large radius chosen such that weakly-bound $B = 0$ states are unaffected. The number of basis functions $N$ is then finite. We use $J_{max} = 36$, $E_{min}/h = -120$ GHz, $E_{man}/h = +60$ GHz, and a largest radius of 52.9 nm. Equation 1 then has $N = 46300$ coefficients. Approximately 7% of the matrix elements of our banded matrix are non-zero and ≈15% of the basis states have $E_\nu^{JM} < 0$. The value of matrix elements between continuum states is a few orders of magnitude smaller than those between bound states.

Our statistical analysis looks at eigenstates $|\Psi; M\rangle$ with energy in the small energy interval $[\mathcal{E}, 0]$ with $\mathcal{E} < 0$ and $|\mathcal{E}| \ll |E_{min}|, E_{max}$. Figure 3 shows these energies for four typical $B$-field strengths. The energy of these eigenstates has converged with respect to basis size. We choose $\mathcal{E}/h = -0.5$ GHz to correspond to the nominal Dy Zeeman energy, $\mu_B B$, at the largest $B$-field strengths studied. Here, $\mu_B$ is the Bohr magneton and $\mu_B/h \approx 1.4$ MHz/G. There are approximately 150 basis functions with energy $E_\nu^{JM} \in [\mathcal{E}, 0]$ corresponding to a mean spacing of $\langle s \rangle/h = 3.3$ MHz. In fact, their nearest-neighbor spacing (NNS) distribution is Poissonian (8). Energy shifts induced by the diagonal, Zeeman matrix elements are already larger than this mean spacing when $B = 1$ G, a field that is smaller than our typical field strength.

We emphasize that the matrix size of the lanthanide Hamiltonian does not play the same role as it typically does in the conventional random-matrix Hamiltonians (32). For random matrices, the eigenenergies do not converge to fixed values with increasing $N$. In fact, the statistical properties of eigenpairs as a function of $N$ is often an integral part of a study and relies on the notion that basis functions are equivalent, i.e. adding spins in a Heisenberg model or lattice sites in an atomic Hubbard model. This is not true for our Hamiltonian as basis states have a distinct physical interpretation. We can increase $J_{max}$ and thus $N$; however, there are diminishing returns from these states, since both the eigenenergies converge and the additional number of zero-field states within the energy range $[E_{min}, E_{max}]$ decreases.

**Acknowledgments:** We thank B. L. Altshuler for an insightful discussion. Funding: Work at the Temple University is supported by the U.S. Air Force Office of Scientific Research and the Army Research Office (grants FA9550-14-1-0321, W911NF-17-1-0563, and W911NF-14-1-0378). Work at the Joint Quantum Institute is supported by the U.S. NSF (grant PHY-1506343).

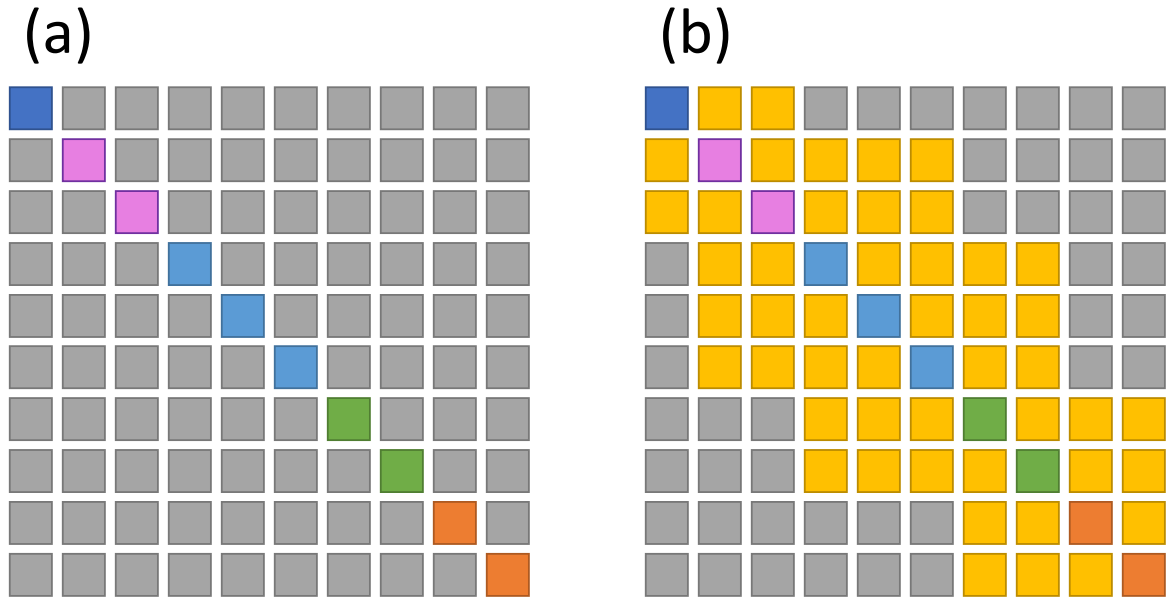

**Fig. 1: Structure of our banded Hamiltonian for lanthanide dimers in a magnetic field with strength *B*.** Grey boxes correspond to matrix elements with a zero value. Others correspond to elements with a non-zero value. **(A)** For B=0 G, the Hamiltonian is diagonal. Boxes with the same color are for states with the same total molecular angular momentum $\vec{J}$. **(B)** For finite $B$, the Hamiltonian is banded. The non-zero off-diagonal matrix elements (yellow boxes) are proportional to $B$.

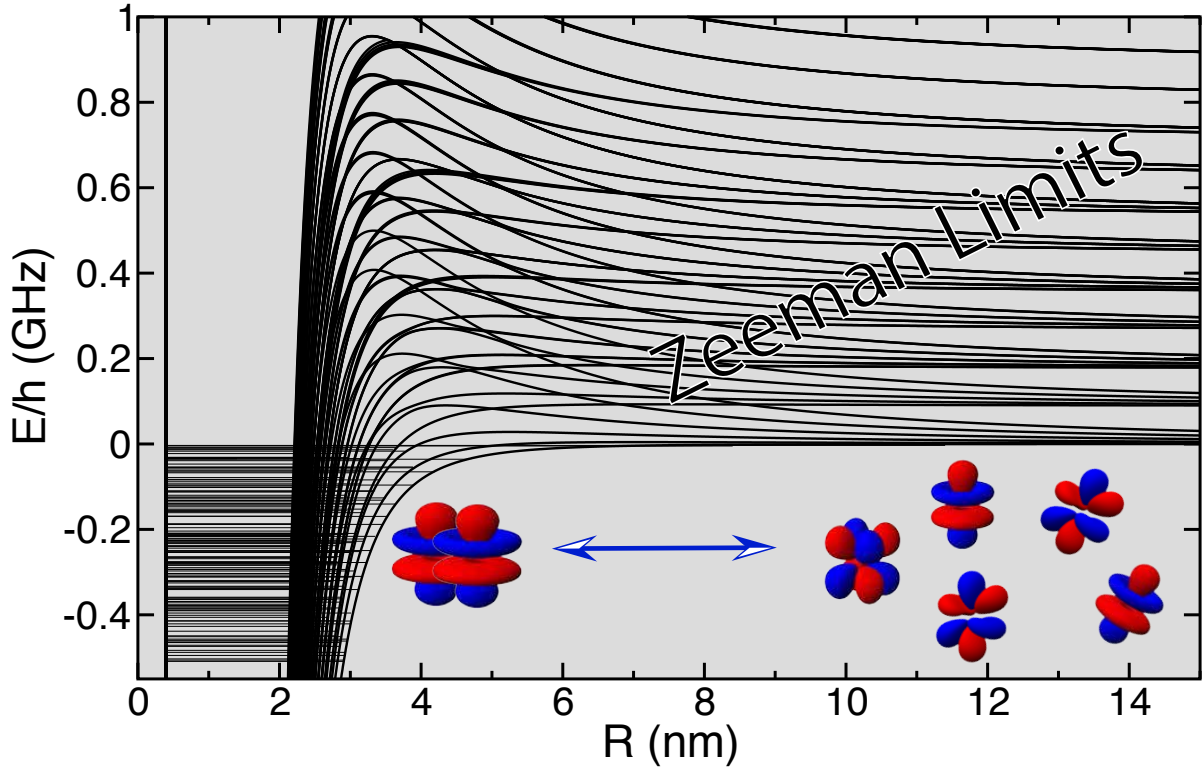

**Fig. 2: Asymptotic thresholds in the $^{164}$Dy+$^{164}$Dy system**. Schematic of (adiabatic) molecular potentials for $B = 50$ G as functions of separation $R$. For graphical purposes, partial waves are restricted to even $\ell$ up to 10. For $R \to \infty$ the potentials dissociate to scattering thresholds labeled by $m_a + m_b = -16, -15, -14, \cdots$ starting from the lowest limit, respectively. Bound states used in the statistical analysis lie within the energy range $\varepsilon/h = -0.5$ GHz and 0 GHz just below the energetically-lowest threshold.

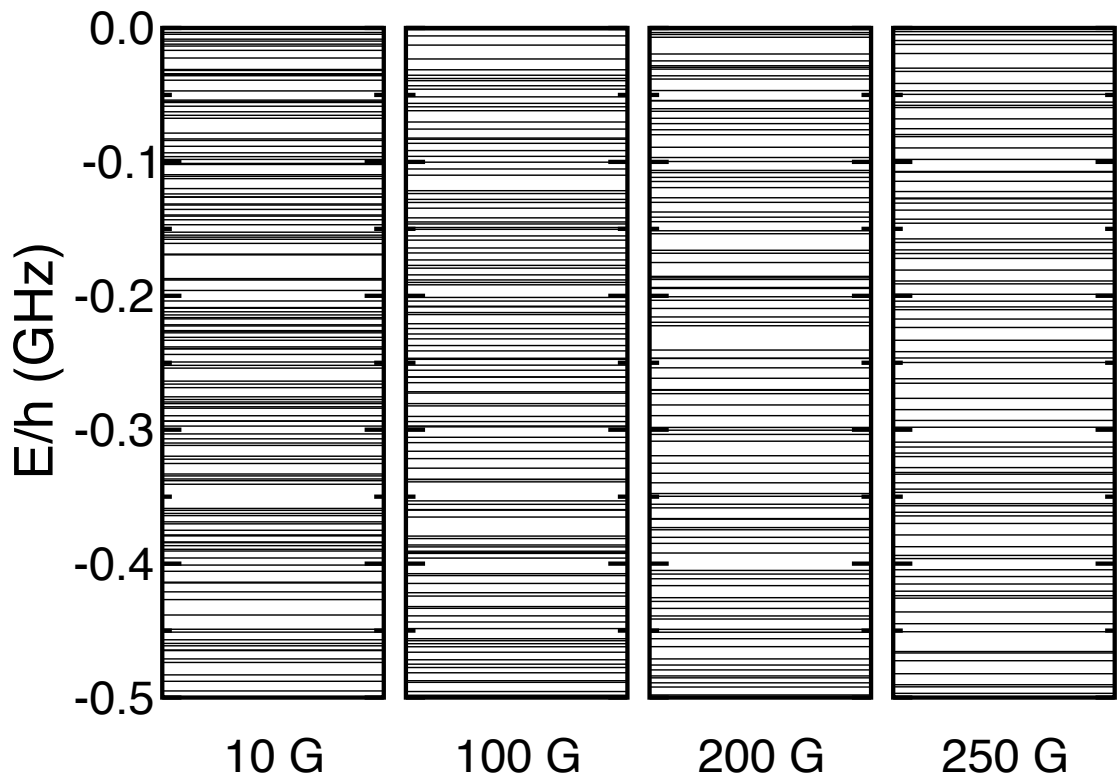

**Fig. 3:** Near threshold eigenstate energies at four typical *B*-field values as indicated below each of the panels.

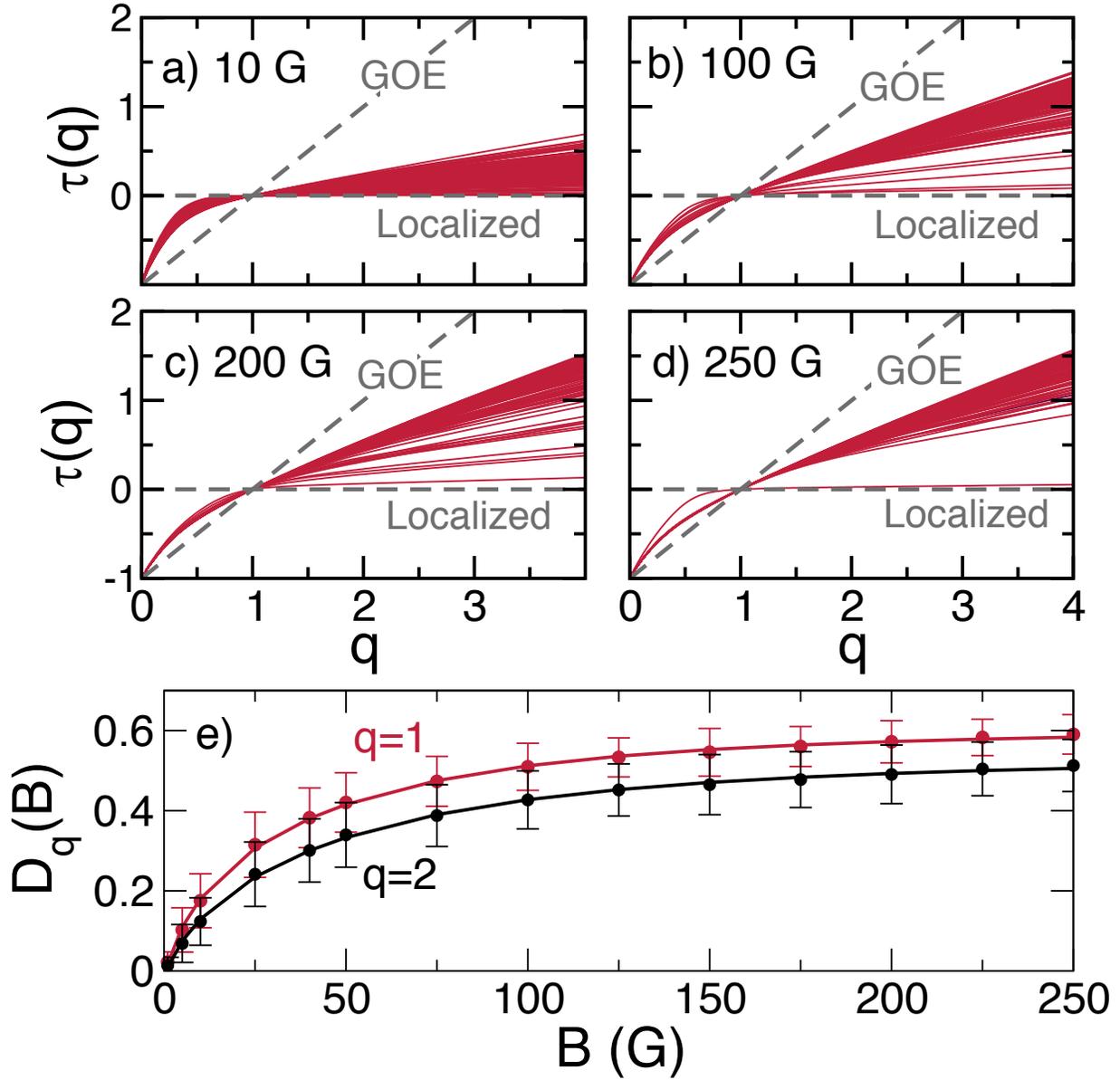

**Fig. 4: Grand potential and fractal dimension of the weakly-bound Dy$_2$ wavefunctions.** (**A** to **D**) show $\tau(q)$ as a function of $q$ for all weakly-bound eigenstates (dark red curves) within a $|\varepsilon/h| = 500$ MHz energy range below threshold at $B = 10$ G, 100 G, 200 G, and 250 G, respectively. The corresponding eigenenergies for these four $B$ fields is shown in Fig. 3. The two dashed straight lines correspond to $\tau(q)$ of the ergodic GOE and Localized limits, respectively. (Although not shown, the curves remain concave for $q < 0$.) (**E**) Fractal dimension $D_q(B)$, defined

in the text, as functions of $B$ for $q = 1$ and 2, respectively. The error bars indicate the spread of fractal dimension. Solid lines are for guiding the eye. For $q = 1$ and 2 the parameter $d_0 = 0.57$ and 0.49, respectively.

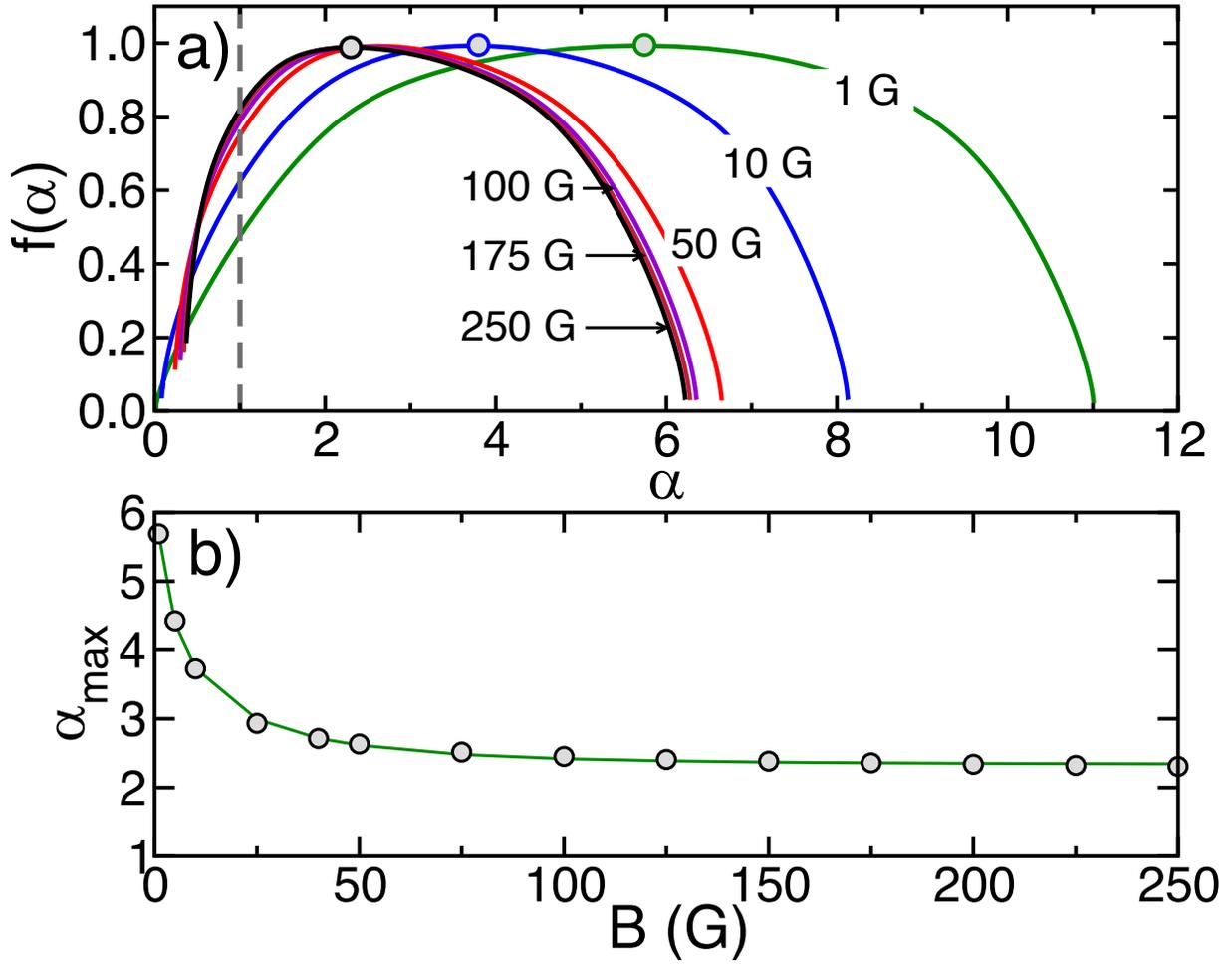

**Fig. 5**: **Multifractality of Dy$_2$ wavefunctions.** (**A**) Spectrum of the fractal dimension $f(\alpha)$ for Dy$_2$ eigenstates within a 500 MHz energy window below threshold as a function of $\alpha$ for $B = 1$ G, 10 G, 50 G, 100 G, 175 G, and 250 G. The dashed vertical line indicates the ergodic limit, where $f(\alpha)$ is a delta function located at $\alpha = 1$. Filled circles for selected $B$ locate the maximum of $f(\alpha)$. (**B**) Location $\alpha_{max}$ of the maximum of $f(\alpha)$ as a function of $B$.

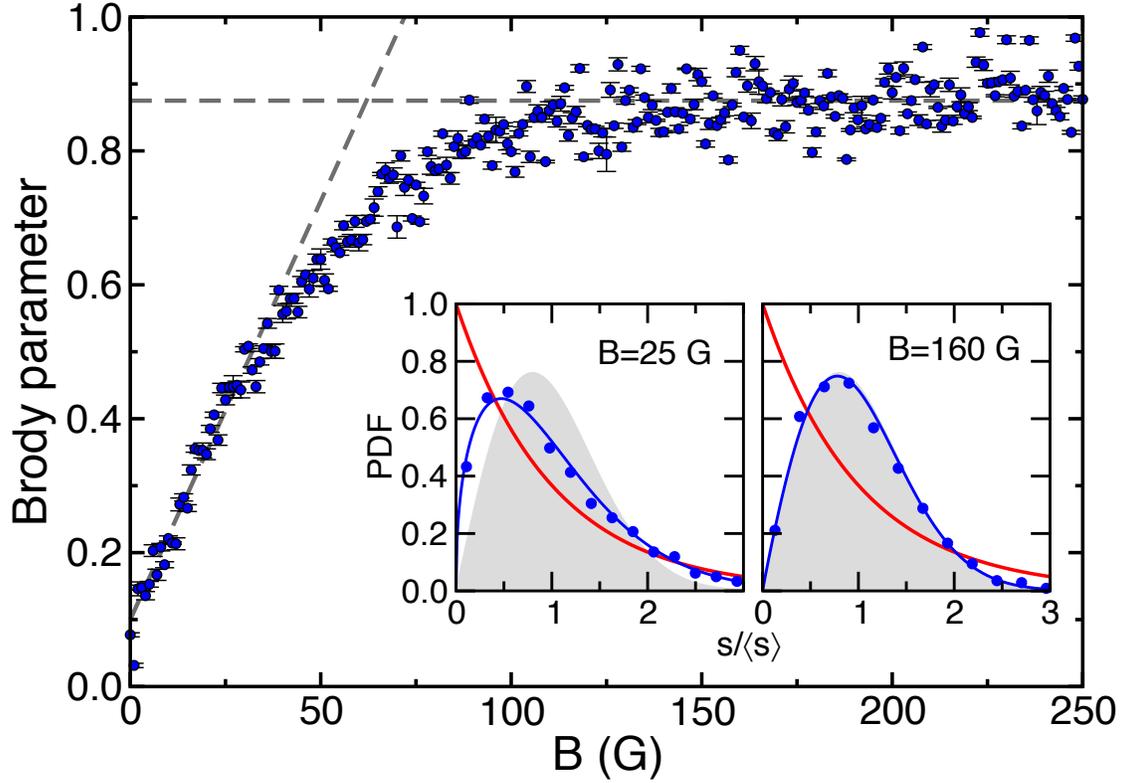

**Fig. 6**: **Brody parameter $\eta$ extracted from the NNS distribution as a function of $B$ for weakly-bound states with binding energy less than $|\varepsilon|/h = 0.5$ GHz**. The dashed gray lines extrapolate the linear small and large $B$ field behavior. Their crossing defines a transition strength. Inserts show the NNS probability distributions (blue filled circles) for $B = 25$ G and 160 G as a function of the normalized energy spacing $s/\langle s \rangle$ with mean spacing $\langle s \rangle$ obtained for each $B$. Solid blue lines are Brody distributions fit to the data. Gray shaded areas and red curves indicate the Wigner-Dyson ($\eta = 1$) and Poisson ($\eta = 0$) distributions, respectively.